\documentclass[twocolumn,showpacs,showkeys,amsmath,amssymb]{revtex4}
\usepackage{times}

\newcommand{\beq}{\begin{equation}}
\newcommand{\eeq}{\end{equation}}

\newcommand{\bfOm}{\mbox{\boldmath $\Omega$}}
\def\bfv{\bf v}

\def\L{\Lambda}

\def\gs{\mathrel{\lower0.6ex\hbox{$\buildrel {\textstyle >}\over{\scriptstyle \sim}$}}}
\def\ls{\mathrel{\lower0.6ex\hbox{$\buildrel {\textstyle <}\over{\scriptstyle \sim}$}}}

\begin{document}

\title{Solar and stellar system tests of the cosmological constant}

\author{Mauro Sereno}
\email{sereno@physik.unizh.ch}

\author{Philippe Jetzer}
\email{jetzer@physik.unizh.ch}

\affiliation{Institut f\"{u}r Theoretische Physik, Universit\"{a}t Z\"{u}rich,
Winterthurerstrasse 190, CH-8057 Z\"{u}rich, Switzerland}

\date{January 16, 2006}

\begin{abstract}

Some tests of gravity theories - periastron shift, geodetic precession, change in mean motion and gravitational redshift - are applied in solar and stellar systems to constrain the cosmological constant. We thus consider a length scale range from $\sim 10^8$ to $\sim 10^{15}~\mathrm{km}$. Best bounds from the solar system come from perihelion advance and change in mean motion of Earth and Mars, $\L \ls 10^{-36}\mathrm{km}^{-2} $. Such a limit falls very short to estimates from observational cosmology analyses but a future experiment performing radio ranging observations of outer planets could improve it by four orders of magnitude. Beyond the solar system, together with future measurements of periastron advance in wide binary pulsars, gravitational redshift of white dwarfs can provide bounds competitive with Mars data.
\end{abstract}


\maketitle

\section{Introduction}

The understanding of the cosmological constant $\L$ is one of the most outstanding topic in theoretical physics. On the observational side, the cosmological constant is motivated only by large scale structure observations as a possible choice for the dark energy \cite{pad05}. In fact, when fixed to the very small value of $\sim 10^{-46}\mathrm{km}^{-2}$, $\L$, together with dark matter, can explain the whole bulk of evidence from cosmological investigations. In principle, the cosmological constant should take part in phenomena on every physical scale but due to its very small size, a local independent detection of its existence is still lacking. Measuring local effects of $\L$ would be a fundamental confirmation and would shed light on its still debated nature, so it is worthwhile to investigate $\L$ at any level. 

Up till now, no convincing method for constraining $\Lambda$ in an Earth's laboratory has been proposed \cite{je+st05}. Astronomical phenomena seem to be more promising. The cosmological constant can affect celestial mechanics and some imprints of $\L$ can influence the motion of massive bodies. In particular, the effect on the perihelion precession of solar system planets has been considered to limit the cosmological constant to $\L \ls 10^{-36}\mathrm{km}^{-2}$ \cite[and references therein]{je+se05}. Not all of the classical tests of general relativity can be applied to constrain $\L$. In fact, the cosmological constant does not participate in the bending of light rays \cite{isl83,lak02}. On the other hand, gravitational time delay of electromagnetic rays is instead influenced \cite{ker+al03,kag+al06}. The cosmological constant could also play a role in the gravitational equilibrium of large astrophysical structures \cite{ba+no05,bal+al06}. On the scale of the Local Volume, $\L$ could have observable consequences by producing lower velocity dispersion around the Hubble flow \cite{tee+al05}.

In this paper, we want to discuss how solar and stellar system observations can be used to give evidence of the cosmological constant. Observations of binary pulsars could be competitive in the near future \cite{je+se05}, but solar system tests are still more effective and continue to provide essential information for undestanding the nature of gravity \cite{ber+al06}. We will focus mainly on planetary perturbations. Since much of the past attention was focused on perihelion shifts \cite{isl83,wri98,ker+al03,je+se05}, here we are more concerned with other alternative proposals. We review and update some previous ideas and discuss some new observational targets. The effect of $\L$ is tested from a length scale of order of $\sim 1~\mathrm{AU}$, by considering the motion of test bodies in bound gravitational systems, to $\sim 10^2~\mathrm{pc}$, by considering the observation of distant white dwarfs. In section~\ref{sec:gyro}, we discuss the effect of the cosmological constant on gyroscope precession. The change in mean motion and the periastron shift are discussed in section~\ref{sec:mean} and in section~\ref{sec:peri}, respectively. Section~\ref{sec:grav} is about what can be obtained from gravitational redshift experiments. Final considerations are contained in section~\ref{sec:conc}.

\section{Gyroscope precession}
\label{sec:gyro}

The effect of $\L$ on the motion of a test body can be considered in the framework of the spherically symmetric Schwarzschild vacuum solution with a cosmological constant, also known as Schwarzschild-de~Sitter or
Kottler space-time \cite{adl+al65}. In the weak field limit, this metric reads
\begin{equation}
\label{gyro1}
ds^2 \simeq (1+2\phi/c^2)(cdt)^2 - (1 - 2\phi/c^2)\delta_{a b}  dx^adx^b,
\end{equation}
where Latin indeces vary over 1,2,3 and the potential $\phi$ is given by
\begin{eqnarray}
\phi & = & \phi_\mathrm{N}     + \phi_\L \\
     & = & - \frac{G M}{r}  - \frac{\L}{6}c^2 r^2.\label{gyro3}
\end{eqnarray}
As can be seen from Eq.~(\ref{gyro3}), in presence of a cosmological constant, there is an upper limit on the
maximum distance within which the weak field limit holds \citep{now01}. For any realistic value of the cosmological constant, the $\L$ contribution to the gravitational potential exceeds the Newtonian one only on a very large scale. For $M \sim M_\odot$ and $\L \sim 10^{-46}~\mathrm{km}^{-2}$, $|\phi_\L| \gs |\phi_\mathrm{N}|$ for $r \gs 150~\mathrm{pc}$. On the other hand, $\phi_\L$ exceeds the weak field limit only on a cosmological scale \citep{now01}. 

According to the general theory of relativity, a spinning gyroscope orbiting around a massive body undergoes precession with respect to the distant standard of rest. For a space-time in the form of Eq.~(\ref{gyro1}), the spin vector precesses due to spin-orbit coupling with the prograde
angular velocity \cite{str04}
\begin{equation}
\label{gyro2}
\bfOm_\mathrm{s-O} = - \frac{3}{2c^2} \bfv {\times}  \nabla \phi .
\end{equation}
where $\bfv$ is the velocity of the body. The main contribution to the precession is the well known de~Sitter or
geodetic precession which, averaged over a revolution, is
\begin{equation}
\label{gyro4}
\bfOm_\mathrm{s-O}^\mathrm{dS} =
\frac{3}{2} \frac{ r_\mathrm{g} }{a^3} \frac{\mathbf{L} }{\left( 1 - e^2 \right)^{3/2}},
\end{equation}
where $\mathbf{L}$ is the specific angular momentum of the unperturbed elliptical orbit, $L =
\sqrt{G M a (1-e^2)}$, $a$ the semi-major axis, $e$ the eccentricity of the orbit of the gyroscope, $M$ the mass of the central body and $r_\mathrm{g} \equiv G M/c^2$ its gravitational radius. Due to $\L$, an additional term appears. The spin-orbit contribution to the precession due to the cosmological constant is
constant and can be written as
\begin{equation}
\label{gyro5}
\bfOm_\mathrm{s-O}^\L =
-\frac{1}{2} \L \mathbf{L} .
\end{equation}
The ratio between the two contributions to the precession is
\begin{equation}
\label{gyro6}
\frac{\bfOm_\mathrm{s-O}^\L}{ \bfOm_\mathrm{s-O}^\mathrm{dS} } 
= -\frac{1}{3} \L \frac{a^3}{ r_\mathrm{g} }\left( 1 - e^2 \right)^{3/2} .
\end{equation}

The orbit of a gravitationally bound system, which is small with respect to the rest of the system, can be used instead of a gyroscope. Analyses of laser ranges to the Moon, via precision measurements of the lunar orbit, have been providing increasingly accurate verification of relativistic gravity, such as equivalence principle violation test and search for a time variation in the gravitational constant \cite{wil+al04}. The Earth-Moon system can be regarded as a gyroscope moving with the Earth, with its axis perpendicular to the orbital plane. In fact, due to spin-orbit precession, the sideral mean motion of the Moon and the lunar perigee and node rates are not changed by the same amount \cite{ber+al87}. An estimate of contribution of geodetic precession to the the lunar perigee demands for an accurate modelling of conventional sources of precession such as Earth's quadrupole field and perturbations from other solar-system bodies. Observed deviations of geodetic precession from its predicted general relativity value of $19.2~\mathrm{mas/year}$ were used to constrain an
Yukawa-like contribution to the gravitational potential \cite{ade+al03}. The analysis of the lunar lase ranging data to April 2004 yielded a relative deviation of geodetic precession from its expected value of $-0.0019 {\pm} 0.0064$ \cite{wil+al04}. The constraint on the cosmological constant from the 1-$\sigma$ lower limit is $\Lambda \ls
1 {\times} 10^{-26}\mathrm{km}^{-2} $.

The Gravity Probe B mission should measure the geodetic precession with an accuracy of about $0.5~\mathrm{mas/year}$. Despite of the high precision, this experiment will be not effective in constraining $\L$. Due to the small orbital radius, the bound on the cosmological constant would be $\Lambda \ls 3 {\times} 10^{-21}\mathrm{km}^{-2} $ \cite{kag+al06}.

Precession of pulsar spin axis due to relativistic spin-orbit coupling
have been recently detected for some binary systems \cite{we+ta02,
hot+al05}, but precision is still relatively low and does not allow to
put any constraint on $\L$.

\section{Mean motion}
\label{sec:mean}

\begin{table}
\caption{\label{tab:mean} Limits on the cosmological constant
due to anomalous mean motion of the solar system planets. $\delta a$ is the statistical error in the orbital semi-major axis \cite{pit05}; $\Lambda_\mathrm{lim}$ is the 1-$\sigma$ upper bound on the cosmological constant.}
\begin{ruledtabular}
\begin{tabular}{lrr}
Name  &  $ \delta a~(\mathrm{km})$ & $\Lambda_\mathrm{lim}~(\mathrm{km}^{-2})$ \\
\hline
Mercury  &  $0.105 \times 10^{-3}$   &  $ 1 {\times} 10^{-34}$
\\
Venus    &  $0.329 \times 10^{-3}$   &  $ 3 {\times} 10^{-35}$
\\
Earth    &  $0.146 \times 10^{-3}$   &  $ 4 {\times} 10^{-36}$
\\
Mars &      $0.657 \times 10^{-3}$   &  $ 3 {\times} 10^{-36}$
\\
Jupiter &   $0.639  \times 10^{+0}$   &  $ 2 {\times} 10^{-35}$
\\
Saturn  &   $0.4222 \times 10^{+1}$   &  $ 1 {\times} 10^{-35}$
\\
Uranus  &   $0.38484\times 10^{+2}$   &  $ 8 {\times} 10^{-36}$
\\
Neptune &   $0.478532\times 10^{+3}$   &  $ 2 {\times} 10^{-35}$
\\
Pluto   &  $0.3463309\times 10^{+4}$  &  $ 4 {\times} 10^{-35}$
\end{tabular}
\end{ruledtabular}
\end{table}

A positive cosmological constant would decrease the effective mass of the Sun as seen by the outer planets. Due to $\L$, the radial motion of a test body around a central mass $M$ is affected by an additional acceleration, $\mathcal{A}_\L
=\L c^2 r/3$,  and a change in the Kepler's third law occurs \cite{wri98}. For a circular orbit,
\begin{eqnarray}
\omega^2 r & =&  \frac{G M}{r^2} - \frac{\L c^2}{3} r  \label{mean1} \\
& \equiv & \frac{G M_\mathrm{eff}}{r^2} . \label{mean2}
\end{eqnarray} where $\omega$ is the angular frequency. By comparing Eqs.~(\ref{mean1},~\ref{mean2}), we get the variation due to $\L$ in the effective mass for test bodies at radius $r$,
\begin{equation}
\label{mean3}
\frac{\delta M_\mathrm{eff}}{M} = - \frac{1}{3}\L \frac{r^3}{r_\mathrm{g}}.
\end{equation}
In other words, the mean motion $n \equiv \sqrt{G M/ a^3}$ is changed by \cite{wri98},
\begin{equation}
\label{mean4}
\frac{\delta n}{n} = - \frac{\L}{6} \frac{a^3}{r_\mathrm{g}}.
\end{equation}
Variation of the effective solar mass felt by the solar system inner planets with respect to the
effective masses felt by outer planets could probe new physics \citep{and+al89,and+al95}. Orbital elements of solar system planets were recently determined with precision EPM ephemerides based on more than 317,000 position observations of different types, including radiometric and optical astrometric observations of spacecraft, planets, and their satellites \cite{pit05}. Ephemerides were constructed by simultaneous numerical integration of the equations of motion in the post-Newtonian approximation accounting for subtle effects such as the influence of 301 large asteroids and of the ring of small asteroids, as well as the solar oblateness. We can then evaluate the statistical error on the mean motion for each major planet, $\delta n = - (3/2) n \delta a/a$, and translate it into an uncertainty on the cosmological constant. Results are listed in Table~\ref{tab:mean}. Best limits comes from Earth and Mars. Errors in Table~\ref{tab:mean} are formal and could be underestimated. Current accuracy can be determined evaluating the discrepancies in different ephemerides \cite{pit05}. Differences in the heliocentric distances do not exceed $10~\mathrm{km}$ for Jupiter and amount to 180, 410, 1200 and 14000~km for Saturn, Uranus, Neptune and Pluto, respectively \cite{pit05}. Bounds on $\L$ from outer planets reported in Table~\ref{tab:mean} should be accordingly increased.

Unlike inner planets, radiotechnical observations of outer planet are still missing and their orbits can not be determined with great accuracy. Apart from optical observations, only Voyager 2 flyby data are available for Uranus and Neptune, with an accuracy in the determination of distance of $\sim 1~\mathrm{km}$ \cite{and+al95}. In fact, the measurements precision of ranging observations is roughly proportional to the range distance. We can assume a conservative uncertainty of $\delta a \sim 10^{-1}$-$1~\mathrm{km}$ on the Neptune or Pluto orbits from future space missions, which would bound the cosmological constant to $\Lambda \ls 10^{-38}$-$10^{-39}~\mathrm{km}^{-2}$, three order of magnitude better than today's constraints from Mars.

Pioneer spacecrafts have been considered as ideal systems to perform precision celestial mechanics experiments \cite{and+al02}. Analyzed data cover a heliocentric distance out to $\sim 70~\mathrm{AU}$ and show an anomalous acceleration directed towards the sun with a magnitude of $\sim 9 {\times} 10^{-8}\mathrm{cm~s}^{-2}$ \cite{and+al02}. If all the systematics were accounted for, that acceleration could be originated by some new physics. An interpretation of these data in terms of $\L$ would imply a negative cosmological constant, which seems quite unlikely. Taken at the face value, the Pioneer anomalous acceleration would give $\Lambda \sim -3 \times 10^{-35}\mathrm{km}^{-2}$.

\section{Perihelion precession}
\label{sec:peri}

The effect of a cosmological constant on the advance of the perihelion has been long investigated as a tool to probe $\L$ on a local scale \cite{je+se05}. The precession angle due to $\L$ after one period is \cite{ker+al03}
\begin{equation}
\Delta \phi_\L = \pi \L \frac{a^3}{ r_\mathrm{g}} (1-e^2)^{1/2} .
\end{equation} 
For small eccentricities, the relative precession rate of the periastron due to $\L$ (i.e. $\Delta \phi_\L/2\pi $) is three times larger than the variation in the mean motion \cite{wri98}. Accurate measurements of Earth and Mars perihelion shift have provided so far the more tight bound on $\L$ from solar system tests, $\Lambda \ls 1 {\times}
10^{-36} \mathrm{km}^{-2}$ \cite{je+se05,ior05}. The precision of a frequency determination can be expressed as $\delta \omega \sim \delta r / (a e t)$, with $\delta r$ being the precision in range and $t$ the time interval for observations \cite{nor00}. The corresponding bound on $\L$ reads
\begin{equation}
\label{peri1}
\L \ls \frac{1}{a^2 e (1-e^2)^{1/2}} \left( \frac{r_\mathrm{g}}{a}\right)^{1/2} \frac{\delta r}{ct}
\end{equation}
The best orbital eccentricity to constrain $\Lambda$ is $e = 1/\sqrt{2} \sim 0.7$, but values in the range $0.54 \ls e \ls 0.84$ are also well suited, with a worsening of less than 10\% with respect to the optimal value. About near future prospects to lower the upper bounds on $\Lambda$, the same considerations done for the mean motion change still apply. Ranging observations would help significantly. Pluto, the planet with both largest eccentricity and major axis, would be the best target for radio observations. On the other hand, satellites orbiting the Earth are too much small systems with respect to planets orbiting the Sun and even very accurate measurements of their orbital elements are not useful in constraining $\Lambda$.

Wide binary pulsar could offer interesting possibilities. For systems such as B0820+02 and J0407+1607, the advance of periastron due to the cosmological constant is $ \sim {\times}10^{27} \Lambda/ (1~\mathrm{km}^{-2}) \deg/\mathrm{days}$, slightly  better than the Mars one \cite{je+se05}.

\section{Gravitational redshift}
\label{sec:grav}

According to general relativity, in presence of a static gravitational field the frequency $\nu$ of a signal transmitted from a clock at rest at $r$ is gravitationally shifted with respect to the frequency measured by an identical standard clock at rest located at a different place ($r_0$) by
\begin{equation}
\label{grav1}
z \equiv \frac{\Delta \nu}{\nu} \simeq \frac{\Delta \phi}{c^2} .
\end{equation}
where $\Delta \phi \equiv \phi (r) -\phi (r_0)$ is the difference in the gravitational potential between the emitter and the receiver. Gravitational redshift experiments have provided crucial tests of the equivalence principle \cite{wil05}. On turn, they could provide accurate measurements of the gravitational potential and, in particular, of the cosmological constant \cite{kag+al06}. 

The solar gravitational redshift was determined by observations of the infrared oxygen triplet both in absorption and in emission. The experimental result was in agreement with the equivalence principle prediction to about 2\% \cite{lop+al91}. From this uncertainty, we get a limit of $\Lambda \ls 10^{-23}~\mathrm{km}^{-2}$. The gravitational redshift effect due to Saturn was tested with an accuracy of 1\% from the Voyager 1 flyby \cite{kri+al90}. Considering the spacecraft periapsis of $1.8 \times 10^5~\mathrm{km}$, we get $\Lambda \ls 7 \times 10^{-28}~\mathrm{km}^{-2}$. A promising gravitational redshift experiment in the outer solar system was proposed  based on a spacecraft equipped with a trapped-ion frequency standard \cite{kri94}. Given a fractional frequency stability of $10^{-17}$, a redshift test could have a detection sensitivity approaching $\Lambda \ls 10^{-37}~\mathrm{km}^{-2}$ for a spacecraft at $\gs 100~\mathrm{AU}$ from the Sun. With respect to the accurate determinations of long orbital periods required by methods discussed in sections~\ref{sec:mean} and \ref{sec:peri}, this experiment could be performed on a much shorter span of time.

Better constraints could be obtained going beyond the solar system. Gravitational redshifts  have been measured for several dozens of white dwarfs \cite{rei96,sil+al01}. When combined with independent estimates of the stellar radius and mass, gravitational redshifts for systems at known distance $d$ could provide interesting constraints on the cosmological constant,
\begin{equation}
\label{grav2}
\L \ls 2 \times  10^{-32}\left( \frac{ c~\delta z}{ 1~\mathrm{km~s}^{-1} }\right) \left( \frac{1 ~\mathrm{pc}}{d}\right)^2~\mathrm{km}^{-2} ,
\end{equation}
where $\delta z$ is the accuracy in the redshift determination. For $c~\delta z  \sim 1~\mathrm{km~s}^{-1}$, as often obtained in observed systems \cite{sil+al01},  and $d \sim 100~\mathrm{pc}$, we get $\L \ls  2 \times 10^{-36}~\mathrm{km}^{-2}$.  The limit from Sirius B, the nearest ($d \sim 2.7~\mathrm{pc}$) and brightest of all white dwarfs, for which  $c~z \sim 80 \pm 5$ \cite{bar+al05}, is $\L \ls 1 \times 10^{-32}~\mathrm{km}^{-2}$. Just as an example of what could be obtained from observations of farther white dwarfs, the gravitational redshift of $c~z = 26.9 \pm 0.9~\mathrm{km~s}^{-1} $ measured for the system 2341-164 at $d \sim 80~\mathrm{pc}$ \cite{rei96} gives an upper bound on the cosmological constant of $\L \ls 3 \times 10^{-36}~\mathrm{km}^{-2}$.

\section{Conclusions}
\label{sec:conc}

We have considered the effect of $\Lambda$ on the precession of a gyroscope, the change in the mean motion and the periastron shift of a massive body and, finally, gravitational redshift. As it could be expected from a dimensional argument, relative variations due to $\L$ always goes as $\propto \L ( a^{3} /r_\mathrm{g}) (a/r_\mathrm{g})^i$, $i=\{0,1,...\}$. Wide system are highly preferred. Using available data of the solar system, the best constraint comes from perihelion precessions of Earth and Mars, $\Lambda \ls 1 {\times} 10^{-36} \mathrm{km}^{-2}$. Analysis of anomalies in the mean motion provide limits at the same order of magnitude, whereas measurements of gyroscope precession and gravitational redshift fall short. Beyond the solar system, similar limits come from gravitational redshift measurements in white dwarfs. Despite non being competitive with the estimate from observational cosmology analyses, $\L \sim 10^{-46} \mathrm{km}^{-2}$, these tests still appear worthwhile to be investigated since they probe the universal origin of the cosmological constant on very different scales. Any detection of perturbations in the orbital motion in a bound gravitational system, either the solar system or a binary pulsar, probes $\L$ on a scale of the order of the astronomical unit. On the other hand, the relevant length scale in measurements of gravitational redshift is the distance to the source, which is of order of $\ls 10^2$~pc for galactic white dwarfs. The experiments we have considered cover a range in distance of nearly seven orders of magnitude, which help in filling the gap between local systems and the cosmological scenario. Near-future technology should allow to improve bounds by several order of magnitude, the crucial step being radio ranging observations of solar system outer planets.

\begin{acknowledgments}
M.S. is supported by the Swiss National Science Foundation and by the Tomalla Foundation.
\end{acknowledgments}

% \bibliography{lambda_test}

\end{document}